# EILEEN: A recommendation system for scientific publications and grants

Daniel E. Acuna, Kartik Nagre[2,+], Priya Matnani[3,+]

[1] School of Information Studies
Syracuse University
Syracuse, NY

[2] Tesla Motor Co.

[3] Airbnb Inc.

**Abstract.** Finding relevant scientific articles is crucial for advancing knowledge. Recommendation systems are helpful for such purposes, although they have only been applied to science recently. This article describes EILEEN (Exploratory Innovator of LitEraturE Networks), a recommendation system for scientific publications and grants with open source code and datasets. We describe EILEEN's architecture for ingesting and processing documents and modeling the recommendation system and keyphrase estimator. Using a unique dataset of log-in user behavior, we validate our recommendation system against Latent Semantic Analysis (LSA) and the standard ranking from Elasticsearch (Lucene scoring). We find that a learning-to-rank with Random Forest achieves an AUC of 0.9, significantly outperforming both baselines. Our results suggest that we can substantially improve science recommendations and learn about scientists' behavior through their search behavior. We make our system available through https://eileen.io

## 1. Introduction

Modern researchers have access to a large number of scientific publications and grants. While this abundance has allowed them to get more information quickly, it has also made it more overwhelming. Many scientists consider the search for related work as a highly time-consuming part of their responsibilities [1]. While modern researchers have access to a wealth of information, we need methods and tools to help navigate them.

Scientists find articles by following references in other articles that they find interesting. This method might lead scientists to citation communities where all articles refer to each other while causing a heavy bias towards already-cited papers [2]. A complementary method to explore articles is using keyword search. While powerful, this approach is limited to finding the keyword that matches the scientist's provided set of keywords. For example, it would be challenging to form queries without knowing the space of available keywords. Furthermore, keyword search does not adapt as keyword definitions might depend on sometimes outdated perceptions of the field.

This paper describes a recommendation system for publications and grants called EILEEN (Exploratory Innovator of LitEraturE Networks). This system uses the behavior of scientists to learn how to improve its predictions. The first part gives an overview of related work and discusses the advantages and disadvantages of existing approaches. The central part introduces its architecture and the approach for modeling keyphrases and recommendations. The last part of the paper gives insights into the search engine and the performance of our recommendation system.

## 2. Background information

### 2.1 Recommendation systems

1. Content-based filtering
2. Collaborative filtering
3. Hybrid systems

2.1.1 Recommendation systems in science

Over the last 16 years, more than 200 articles have been published about research paper recommendation systems. Among them, 55% of the system applied content-based filtering, while collaborative filtering was applied by only 18%, and graph-based recommendations by 16% [3]. Citation databases such as CiteSeer apply citation analysis to identify articles that are similar to the input article. Scholarly search engines such as Google Scholar focus on classic text mining and citation counts [4]. However, each concept has disadvantages, such as focusing on already-cited articles or showing only limited sources. All these disadvantages limit their suitability for generating recommendations.

### 2.2 Keyphrase prediction

Keyphrase extraction is the task of representing the main content of a document with keywords. Keyphrases or keywords are a collection of words about the document that best describes the document's nature and core content. Document summarization, categorization, and clustering are tasks where keyphrase extraction is essential [20]. There are several unsupervised learning approaches to extracting keyphrases from the document, such as Text Rank algorithm, RAKE (Rapid Automatic Keyphrase Extraction) algorithm, and KERT (Keyphrase Extraction and Ranking by Topic). We now review them.

*The TextRank algorithm* is a graph-based ranking model for text processing [16]. In a graph-based keyphrase extraction algorithm, the ranking of the keyphrase is determined by the importance of the vertex. Vertices cast a "Vote" or "Rank" other connected vertices. The number of votes cast to a vertex is directly proportional to the importance and rank of the vertex (TextRank). Graphs are built to represent text, depending on the application at hand. Text units of various sizes and characteristics can be added as vertices in the graph, e.g., words, collocations, entire sentences.

*Rapid automatic keyphrase extraction (RAKE)* is an approach introduced in 2010 [17]. It depends on four steps: candidate selection, properties calculation, scoring, and word selection. RAKE works by letting users select the criteria to extract keyphrases. The criteria include selecting the minimum number of characters in each word, the maximum number of words in each phrase, and the minimum number of times the key phrase appears in the document. In addition, punctuation marks are treated as sentence boundaries, and all the words in the stopwords list are also treated as sentence boundaries. Finally, after the candidates are selected, they are scored by their frequency and length. Among the few weaknesses of RAKE are its limited accuracy and inability to normalize the keyphrases using stemming and lemmatization.

*Keyphrase extraction and Ranking by topic (KERT)* is an approach that does not follow the traditional approach of *n*-gram tokenization and ranking system [18]. Instead, KERT constructs topical phrases immediately after clustering the unigrams. KERT goes through three steps in extracting keyphrases. Step one includes clustering words in the document dataset into several foreground topics and one background topic, using background LDA [24]. Step two includes extracting candidate keyphrases for each topic according to the word topic assignment. Finally, step three includes ranking the keyphrases in each foreground topic by integrating the criteria several criteria to rank the results.

## 3 EILEEN: goals and system architecture

We developed EILEEN[1] to help junior faculty and researchers find the right people, grant programs, and journals similar to their interests. The ultimate goal is to serve as an open, research-driven platform that produces 1. curated datasets from a combination of sources, 2. data about scientists' system usage, and 3. results of experiments run for a different combination of users. Currently, the systems that offer recommendations such as Semantic Scholar, Microsoft Academic, and Google Scholar are not open. It is hard or impossible to access their data (except Microsoft Academic through the Microsoft Academic Graph), the user's behavior, and run experiments on them. It is the hope that EILEEN fills this gap in recommendation systems research for science. EILEEN is provided through a web interface available at https://eileen.io. It aims to have a user-friendly and responsive interface. EILEEN records all the actions that scientists take on the website. For example, we know what they preferred in the past and their current search; therefore, we can predict which documents will be found relevant. The overall architecture of EILEEN contains data ingestion, data processing, modeling, and the back and front end (Fig. 1).

*Data ingestion*. The data ingestion and processing run automatically using an Apache Airflow job. Currently, we refresh the data weekly. The data processing is done with an Apache Spark cluster with ten nodes, a total of 768 GB of RAM, 40 TB of disk space, running a Hadoop cluster.

The entire data source comprises 28 million records (numbers from 2018), of which 25 million are extracted from MEDLINE, 2.6 million grants from Federal Exporter, 1 million from Pubmed Open Access Subset, 0.6 million for Arxiv, and 14K from the National Bureau of Economic Research [NBER]. Apache Spark is used with Python to pull the data from the above sources. The raw data obtained from the third-party sources is cleaned, processed using Spark's DataFrame and Machine Learning APIs. Finally, the relevant and required fields are extracted from all data sources, and a consolidated model is created with is later exported to the Elasticsearch index. The model contains the following fields along with the description:

1. ID: internal ID of the document
2. Source: the source of the document (PubMed, arXiv, etc.)
3. Source_id: source-specific ID of the document
4. Type: publication/grant
5. Title: Title of the publication or grant
6. Venue: Venue where the publication was published (e.g., journal or conference name)
7. Abstract: Abstract of the publication or grant summary
8. Scientists: authors of the publication or grant
9. Organizations: affiliations of authors in publication or grant
10. Date: date the publication published or grant awarded
11. Content:
12. End_Date: Grant's end date
13. City: City of the publication
14. Country: Country of the publication
15. OtherID: in case the document has multiple identifiers such as DOI or PMID
16. Tfidf: TFIDF vector for the document
17. Topic: Latent Semantic Analysis of the TFIDF vector
18. Topic_Norm: Topic vector, normalized
19. Buckets: Locality-sensitive hashing of the Topic_norm vector, used for fast recommendations

---

[1] The name EILEEN was created in dedication to a retiree grant administrator in the School of Information Studies at Syracuse University named Eileen Allen (http://www.eileenalleneditorial.com), who in many ways provided similar help to faculty of the school by finding suitable collaborators and grant opportunities.

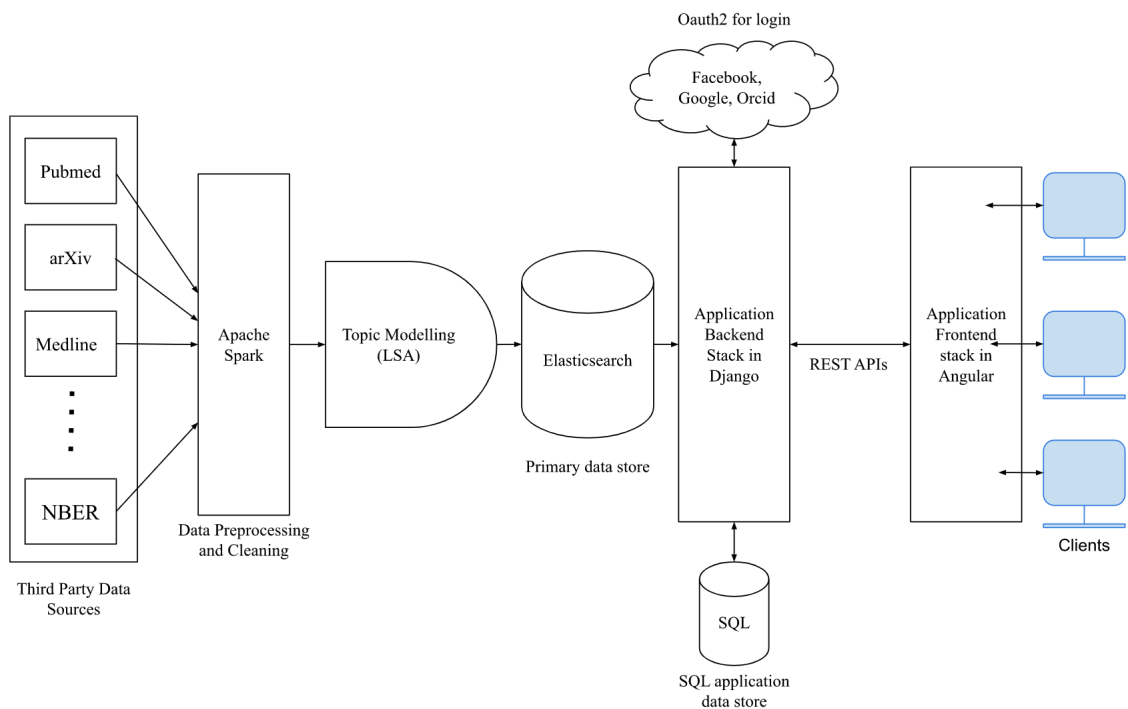

**Fig. 1. EILEEN architecture.** From left to right: multiple sources of documents, including PUBMED, arXiv, and NBER, are downloaded, processed, and homogenized with Apache Spark into a single Parquet table file. The title and abstract of these documents (which might correspond to abstracts and summaries in publications and grants, respectively) are analyzed with Latent Semantic Analysis based on the TF-IDF representation. All the information from the documents is then inserted in an ElasticSearch instance. This instance is used by the backend of our system based on Django and an SQL server to store users' preferences and information. The front end is based on AngularJS. Clients interact with EILEEN through this front-end. For example, if the user wants to register and log in, the request is forwarded to the backend, which does the Oauthr2 authentication with several providers, including Google, Twitter, and ORCID.

The data is dumped periodically from Apache Spark to EILEEN's Elasticsearch server [22].

*Back and front end.* The front end uses AngularJS, an open-source front-end web framework. The user has access to all the EILEEN APIs. We also have an SQL database, which is used for maintaining user data. All the user actions are recorded in the database, including logging in, adding publications and grants to the user's library, and tracking the user's activities. Oauth2 is used to implement login for Google, Facebook, and Orcid APIs.

### 3.1 Functionality

*Keyphrase and search.* Keyphrases are words that capture the main topics of a document. The key phrase extraction system serves two purposes. First, they are displayed on the keyphrase popularity chart in the user library section, extracting relevant key phrases from the user's voted publications and grants. Their popularity is shown across different years. This popularity chart gives users the ability to understand where their interests lie more broadly. Second, they are used to assist the autocomplete function in EILEEN's search functionality.

*Recommendation system.* Once users have chosen a set of publications or grants that interest them, they can use the recommendation system to explore which other publications and grants are similar to their interests. We will explain this part in the methods section.

### 4. Datasets and methods for predicting scientists' behavior

### 4.1 Datasets

#### *4.1.1 Publication and grant datasets*

The datasets for the recommendation system are scientific publications and grants from online repositories such as arXiv, MEDLINE, and Federal ExPORTER.

- arXiv has e-prints of scientific papers in fields such as mathematics, physics, astronomy, computer science, quantitative biology, statistics, and quantitative finance. We download 1.8M scientific papers from arXiv through its API.
- MEDLINE contains citations and abstracts for biomedical literature. Some of them have links to full-text articles. The website has academic journals covering medicine, nursing, pharmacy, dentistry, veterinary medicine, and health care. It also has literature in biology and biochemistry, as well as fields such as molecular evolution. In total, we download around 28M articles from more than 8K journals from MEDLINE.
- Federal ExPORTER provides information about the projects, abstracts, and publications for different federal agencies. We downloaded 2.4M past grants from 14 federal agencies:
  - Administration for Children and Families (ACF)

- Agency for Healthcare Research and Quality (AHRQ)
- Agriculture Research Service (ARS)
- Center for Neuroscience and Regenerative Medicine (CNRM)
- Centers for Disease Control and Prevention (CDC)
- Congressionally Directed Medical Research Programs (CDMRP)
- Defense and Veterans Brain Injury Center (DVBIC)
- Environmental Protection Agency (EPA)
- Food and Drug Administration (FDA)
- Forest Service (FS)
- National Aeronautics and Space Administration (NASA)
- National Institute of Food and Agriculture (NIFA)
- National Institutes of Health (NIH)
- National Institute on Disability, Independent Living, and Rehabilitation Research (NIDILRR)
- National Science Foundation (NSF)
- U.S. Department of Veterans Affairs (VA)

*4.1.2 Behavioral data from EILEEN*

The data captured by eileen.io contains many valuable behavioral items. Of all the users who have visited the website, 28% created a profile and accessed recommendations based on their preferences. As of September 2018, which is the data we use for this study, we have comprehensive data from 68 anonymized users tracked over many months—a high-quality subset of the total. These users spend an average of 2 minutes and 32 seconds on the website, using the search functionality 9.7 times, and saving 14.6 items in their profile.

To build a dataset for a recommendation system, we need to extract several interactions for each user from the backend. We first need to extract the search behavior and user preference from the tracking data to do this. We then find the relationship between the documents searched by the user and the user's votes (i.e., relevant or irrelevant). Then, along with the search result from the Elasticsearch on each user query input, we can build the relationship between users. First, we go through the tracking data to get the queries for the search and the preference on the user's library in the sequence of time. Based on this, we can get all the documents Elasticsearch will return for each query users search for. This process creates three types of preferences:

1. Those users get from the search and vote as 'relevant,'
2. Those users get from the search and do not vote on; we may think those as not so related to the topic users search for,
3. Those users got from the search and voted as 'irrelevant,' which users do not want to have in the search result.

From the result of Elasticsearch for each document, we will have the topics represented for this document in the documents set. So for each user's library, we can calculate the average vector based on all the relevant documents the user voted for. It will represent the topics for this user's library. These steps will allow us to have a dataset for learning a recommendation system for scientists.

**4.2 Methods**

*4.2.1 Content-based recommendation of scientific documents*

The recommendation system we are using is based on the content of a document instead of applying collaborative filters to the data. Thus, we prevent Matthew's effect by avoiding collaborative filters, commonly found in most recommendation systems, which can be an obstacle to scientific exploration [4]. Another feature of our recommendation system is that the users can get feedback immediately after changing their preference by marking one document as relevant or irrelevant. Our recommendation system is seeded by a Rocchio algorithm, extensively used in other scientific recommendation systems [5]. To build this recommendation system, we have to transform the text into vectors, perform dimensionality reduction, create a nearest-neighbor search for initial recommendations, and a learning-to-rank method for re-ranking. We explain these steps now:

*Text preprocessing and term weighting schemes*. The text preprocessing technology we used on our documents set includes removing stop words, stemming with the Porter Stemming algorithm [5], tokenizing the documents into unigrams and bigrams, and filtering it based on term frequency and tf-idf. In particular, we removed the terms that appear fewer than three times or terms whose tf-idf was more than 0.8. We will explain tf-idf now.

There are many different weighting schemes from the researches. They all have different considerations on balancing between the local structure of each document and the global structure of the whole document set. We use this simple weighting scheme called term frequency-inverse document frequency (tf-idf) for our text processing.

The term frequency is based on a matrix where $f_{ij}$ is the frequency of the term $i$ in document $j$. And the tf-idf would reweight term frequency by a global factor as follows

$$tfidf_{ij} = (1 + \log f_{ij}) * \log \frac{N}{f_i + 1}$$

where $f_i$ is the number of documents that contain the term $i$ and $n$ is the total number of documents.

*Latent Semantic Analysis*. Latent semantic analysis (LSA) analyzes relationships between a set of documents and the terms they contain. LSA assumes that words that are close in meaning will occur in similar pieces of text (the distributional hypothesis). A matrix containing word counts per document (rows represent unique words and columns represent each document) is constructed from an extensive collection of documents. A mathematical technique called singular value decomposition (SVD) is used to reduce the dimensionality of the data.

*Rocchio Algorithm*. The Rocchio algorithm provides recommendations based on users' preferences on the documents, and their votes are either relevant or irrelevant [6]. Given a set of relevant documents which some specific user voted previously $\{r_1, r_2, ..., r_n\}$ and another set of irrelevant documents $\{u_1, u_2, ..., u_n\}$, the Rocchio algorithm defines a document vector as follows:

$$Q_m = (a \cdot Q_o) + (b \cdot \frac{1}{|D_r|} \cdot \sum_{D_j \in D_r} D_j) - (c \cdot \frac{1}{|D_{nr}|} \cdot \sum_{D_k \in D_{nr}} D_k),$$

where $Q_0$ is the mean document vector of the users' search on the keywords. Parameters $b$ and $c$ present the weight of the relevant documents set and the irrelevant documents set for the final query vector. For our project, we calculate the search on the keywords into the final document vector; therefore, our query vector is as follows:

$$Q_m = (a \cdot Q_o) + (\frac{1}{|D_r|} \cdot \sum_{D_j \in D_r} D_j)$$

Our recommendation system cares more about the documents that users are interested in than those they show no interest in. So we set the $b$ to be one and the $c$ to be 0.

Once the preference vector $q$ has been created for a user, we will rank the recommendation result from Elasticsearch by the distance to the query vector. Finally, we compute the cosine distance between the word vector representing each recommended document and the query vector in our implementation.

*Learning-to-rank with Random Forest*

Random Forest is an ensemble learning method for classification, regression, and other tasks [7]. And we can use it to rank the importance of variables in a regression or classification problem [8].

After we have all data prepared, we can do the Random Forest classification on all the documents returned from users' search and predict which is more likely to be clicked. We will use the result from the first page only, which means at most ten documents per search. Since we only care about what users are interested in, we will have two classes, one for those documents voted as relevant, and one for those documents remaining un-voted from the search.

We compute twelve features in total. Five of them are for the document, including length of the title, authors, venue, abstract, and search query length. Another set of features is the year of the document and the score returned by Elasticsearch. This Elasticsearch scoring is based on BM25 and other details inherited from the Lucene indexing method[2]. For the rest, we fit the tf-idf on the combined text from query, title, authors, venue, and abstract and then transform the model on each of these five fields. Then we calculate the cosine distance between the tf-idf of the query and the tf-idf of the rest of the fields. This calculation gives us four features. The last feature is the cosine distance between the topics for the document and those of the user's library. This distance is calculated differently based on the document's class. Relevance is

---

[2] https://lucene.apache.org/core/3_5_0/scoring.html

calculated as the cosine distance between the current document and the user's preferences. And for the un-voted documents, it is the cosine distance between this document and the library's state. To summarize, the total set of features is

1. Length of the query
2. Length of the title
3. Length of the authors
4. Length of the venue
5. Length of the abstract
6. Year
7. Score from Elasticsearch
8. Cosine between query and title
9. Cosine between query and authors
10. Cosine between query and venue
11. Cosine between query and abstract
12. Cosine between document and library, which stands for user preference

We use the `RandomForestClassifier` from PySpark with 500 trees. Based on the size of our data, the number of total entries we have is 1302. We randomly split the user in a 4:1 ratio for test and validation. We build five models based on the same data set:

1. All 12 features
2. All 12 features without the score from Elasticsearch
3. All 12 features without the cosine distance between each document and the library
4. Score from Elasticsearch only, and
5. Score from Elastic search and the cosine distance between the document and library

*4.2.2 Keyphrase methods*

The outline of the technique uses a standard algorithm to generate candidate keyword lists and then a model to rerank this initial set of candidates.

*The initial set of candidates.* The algorithm RAKE works in three steps. First, all possible keywords are extracted from the document. Second, a set of properties is calculated to determine the score of each candidate. Third, candidates with top scores (e.g., top 10) are selected. With RAKE, every keyphrase should have five words in length, and it should occur at least three times across the document. For every keyphrase found using RAKE, we include the feature term count, term length, maximum word length, spread, lexical cohesion, and absolute first and last occurrence.

*Training process*. We use a Random Forest using the features from RAKE to learn what constitutes good key phrases or not. Data from [19] is used as the training data for the model that re-ranks the initial set provided by RAKE. It provides 244 scientific documents, out of which 100 are test data set, 144 are training data set, and 40 are validation data set.

# 5. Results

## 5.1 Recommendation system results

As a baseline, we estimated the performance of simple Elasticsearch based on the default weighting of terms in the search query (Fig. 2, left panel). The AUC of such a model is 0.53. By adding the preference vector representation to the model and measuring the cosine distance of such vector with the document vector, the model increases performance significantly (AUC = 0.66). This baseline will be used for comparison.

When testing the performance of all the features described in the Materials and Methods section, we found a performance of AUC = 0.901. We further checked whether the semantic representation of the documents helped this model, but we found that the performance was very close (AUC = 0.903). This performance may indicate that the initial keyword matching captures most documents. These results show that the full model is significantly better than keyword weighting schemes.

With Random Forest, we can interpret which features are most important for scientists when choosing relevant documents (Fig. 2, right panel). For example, we found that for scientists, the abstracts of the documents are more important than titles. In contrast, information retrieval for the web (i.e., normal users) usually finds the essential information piece in the title. Importantly, we found that the score from Elasticsearch is only the 6th most crucial feature.

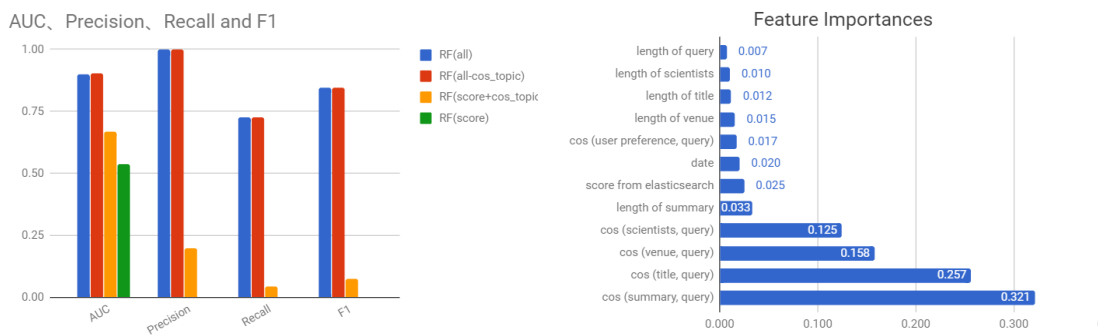

**Fig 2. Comparison of algorithms and feature importance**. (left panel) The Random Forest model with all features (RF(all)) has the best overall performance. (right panel) the features importances of the best model signal that the similarity between the publication abstract or grant summary and the query string is the most important predictor of what the users will find relevant.

## 5.2 Keyphrase extraction results

### 5.2.1 Keyphrase extraction evaluation

The keyphrases generated by RAKE are matched with the extracted keywords provided by semEval. In the documents we are analyzing, we have millions of candidate keywords (Table 1). The intersection of all the keyphrases from RAKE and semEval is 1, and non-matched is 0. The training set of 144 research articles with the condition that keyphrases should be at least five words in length and occur three times across the document produces 1,122 RAKE keyphrases. Out of these, 119 are labeled 1 (Table 1). We ran the Random Forest classifier on the trained data. Cross-validation is performed to find the model with the best binary classification evaluation performance. An AUC = 0.73 is achieved (Fig. 3)

| Totals keyphrases extracted using RAKE and Textrank (Size of training data) | 18,821,440 |
|---|---|
| Keyphrases with label 1 (Matching with human-curated keyphrases) | 1,850,534 |
| Keyphrases with label 0 (Not Matching with human-curated keyphrases) | 16,970,906 |

**Table 1.** Statistics of training data used for EILEEN

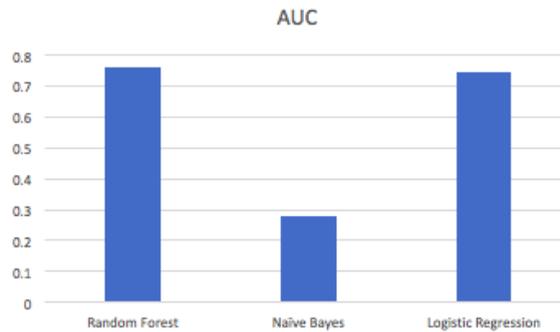

**Fig. 3. Keyphrase extraction model trained on Pubmed Open Access abstracts and semEval prediction.** The Random Forest model has the best overall performance in terms of Area Under the ROC curve.

*Keyphrase prediction for MEDLINE.* A stop word list file, which consists of around 550 stopwords, filters the stopwords. The minimum length of each keyword is 3, the maximum number of keywords is 4, and the minimum occurrences of the phrase are 2. Out of 5,896,505 key phrases from the MEDLINE data, only 40,005 are keyphrases over four keywords, which account for only 0.67% of all human-curated keyphrases. Hence those are filtered out, and the maximum number of keywords in one phrase is set to 4.

| Keywords summary | Counts |
|---|---|
| Number of documents | 2,129,524 |
| Total keyphrases from abstract and title | 5,896,505 |
| Avg keyphrases per document | 2.768 |
| Keyphrases with one word | 3,183,302 |
| Keyphrases with two words | 1,982,907 |
| Keyphrases with three words | 563,094 |
| Keyphrases with four words | 127,197 |
| Keyphrases with five words | 27,324 |
| Keyphrases with more than five words | 12,681 |

**Table 2.** Statistics of the keywords in EILEEN.

Examples of keyphrases produced by our model:

- **Title:** Ecological change as a factor in renewed malaria transmission in an eradicated area. A localized outbreak of A. Aquasalis-transmitted malaria on the demerara river estuary, British Guiana, in the fifteenth year of A. Darlingi.

- **Human Curated Keywords:** Anopheles, Malaria, British Guiana
    **Matched using RAKE and Textrank:** Malaria, British Guiana
- **Title:** Quantitative studies of the effect of organic substrates and 2,4-dinitrophenol on heterotrophic carbon dioxide fixation in hydrogenomonas facilis
    **Human Curated Keywords:** Carbon dioxide, ribose, glucose
    **Matched using RAKE and Textrank:** ribose, glucose

See Appendix section A.1 for other examples.

## 6. Discussion and conclusions

This article analyzed a unique dataset of scientists' behavior when searching for publications and grants. We found that the distance of summary to the query is more important than to title, which seems contrary to regular users of the web. Furthermore, we found that a model with the scientists' preferences and relationships between search terms and the document produced a surprisingly high AUC equal to 0.9. Our analysis shows that we can learn a great deal about scientists' search behavior with only limited data (compared to other systems such as Google Scholar or Semantic Scholar).

One of the issues of this study is that the data is somewhat limited. For example, a typical information retrieval dataset might have thousands of user interactions [23]. Therefore, it could be the case that once we have more data about scientists, the differences found by our model will start to disappear. Also, we do not have impact measures, such as citations or paper downloads, which would better inform our algorithm. However, impact measures are hard to obtain because indexing companies like Elsevier and Thomsons Reuters are usually under intellectual protection. In the future, we will match the documents from MEDLINE to recently released datasets of citations from Microsoft Academic Graph. Still, our findings could inform future research into recommendation systems for scientists.

Another limitation of our study is that the recommendation system only works based on individuals, not using collective behavior. For example, we could imagine an improved system that learns faster about users by combining individual preferences with the preferences of similar scientists. These systems are called hybrid recommendation systems, combining collaborative filtering and content-based recommendation systems. We plan to explore this possibility in the future.

This work introduces several improvements compared to previous work. First, our dataset is unique as it is the only one about scientists' behavior using recommendation and search systems. Second, we also make the code to run the analysis openly available so that the scientific community tests other approaches and compares them to our work. With this, we hope to make research in scientific recommendation systems more reproducible and robust. Finally, we make the results of our best algorithm available through http://eileen.io. Thus our proposal introduced several significant improvements to scientific recommendation systems.

Modern science depends on critically examining the most relevant and recent developments. Recommendation systems should play an increasing role in helping through this examination.

In addition, these new developments could potentially spur faster development of innovations which could improve technologies.

**Conflict of interest statement**

Kartik Nagre is currently affiliated with Tesla Motor Co. and Priya Matnani is currently affiliated with Airbnb Inc. However, both authors performed the work while affiliated with Syracuse University. The authors declare that the research was conducted in the absence of any commercial or financial relationships that could be construed as a potential conflict of interest

**Acknowledgments**

DEA was supported by NSF grants #1646763 and #1800956.

# Appendix

## A.1 Example documents

*Document #1*

*Title*: Exposure to vector-borne pathogens in candidate blood donor and free-roaming dogs of northeast Italy.

*Abstract*: Many vector-borne pathogens including viruses, bacteria, protozoa and nematodes occur in northeast Italy, representing a potential threat to animal and human populations. Little information is available on the circulation of the above vector-borne pathogens in dogs. This work aims to (i) assess exposure to and circulation of pathogens transmitted to dogs in northeast Italy by ticks, sandflies, and mosquitoes, and (ii) drive blood donor screening at the newly established canine blood bank of the Istituto Zooprofilattico Sperimentale delle Venezie. Blood samples from 150 privately-owned canine candidate blood donors and 338 free-roaming dogs were screened by serology (IFA for Leishmania infantum, Ehrlichia canis, Anaplasma phagocythophilum, Babesia canis, Rickettsia conorii, R. rickettsii), microscopic blood smear examination, and blood filtration for Dirofilaria spp. All candidate donors and seropositive free-roaming dogs were tested by PCR for L. infantum, E. canis, A. phagocythophilum, Babesia/Theileria and Rickettsia spp. The dogs had no clinical signs at the time of sampling. Overall, 40 candidate donors (26.7 %) and 108 free-roaming dogs (32 %) were seroreactive to at least one vector-borne pathogen. Seroprevalence in candidate donors vs free-roaming dogs was: Leishmania infantum 6.7 vs 7.1 %; Anaplasma phagocytophilum 4.7 vs 3.3 %; Babesia canis 1.3 vs 2.7 %; Ehrlichia canis none vs 0.9 %; Rickettsia conorii 16 vs 21.3 % and R. rickettsii 11 vs 14.3 %. Seroreactivity to R. rickettsii, which is not reported in Italy, is likely a cross-reaction with other rickettsiae. Filariae, as Dirofilaria immitis (n = 19) and D. repens (n = 2), were identified in free-roaming dogs only. No significant differences were observed between candidate donors and free-roaming dogs either in the overall seroprevalence of vector-borne pathogens or for each individual pathogen. All PCRs and smears performed on blood were negative. This study demonstrated that dogs are considerably exposed to vector-borne pathogens in northeast Italy. Although the dog owners reported regularly using ectoparasiticides against fleas and ticks, their dogs had similar exposure to vector-borne pathogens as free-roaming dogs. This prompts the need to improve owner education on the use of insecticidal and repellent compounds in order to reduce the risk of arthropod bites and exposure to vector-borne pathogens. Based on the absence of pathogens circulating in the blood of healthy dogs, the risk of transmission of these pathogens by blood transfusion seems to be low, depending also on the sensitivity of the tests used for screening.

*Predictions for document #1*

```
+--------+--------------------+---------------------------------------+--------------+-----+
|document|keyword             |probability                            |probability_tf|label|
+--------+--------------------+---------------------------------------+--------------+-----+
|27357128|blood               |[0.8135103281918694,0.18648967180813067]|false        |0    |
|27357128|borne pathogens     |[0.5897005635812653,0.4102994364187347] |true         |0    |
|27357128|candidate blood     |[0.8285770054233941,0.1714229945766058] |false        |0    |
|27357128|exposure            |[0.868695482821145,0.13130451717885508] |false        |1    |
|27357128|free-roaming dogs   |[0.6784501324948651,0.321549867505135]  |true         |1    |
|27357128|italy               |[0.8947328896125741,0.10526711038742595]|false        |1    |
|27357128|northeast           |[0.8204488139426649,0.17955118605733508]|false        |0    |
|27357128|roaming dogs        |[0.6099656603493449,0.39003433965065515]|true         |0    |
|27357128|vector-borne        |[0.723498120350724,0.2765018796492759]  |true         |0    |
|27357128|vector-borne pathogens|[0.594710539721726,0.40528946027827406]|true         |1    |
+--------+--------------------+---------------------------------------+--------------+-----+
```

*Document #2*

*Title*: Different expression profiles of Interleukin 11 (IL-11), Intelectin (ITLN) and Purine nucleoside phosphorylase 5a (PNP 5a) in crucian carp (Carassius auratus gibelio) in response to Cyprinid herpesvirus 2 and Aeromonas hydrophila.

*Abstract*: Interleukin 11 (IL-11), Intelectin (ITLN) and Purine nucleoside phosphorylase 5a (PNP5a) play important roles in innate immunity. In a previous study to identify differentially expressed immune-related genes, suppression subtractive hybridization (SSH) assay was used to characterize differentially expressed genes in crucian carp (Carassius auratus gibelio) infected with Cyprinid herpesvirus 2 (CyHV-2) in which IL-11, ITLN and PNP5a were identified to be the three most significantly up-regulated genes (Xu et al., Archives of Virology, 2014, http://dx.doi.org/10.1007/s00705-014-2011-9). In this study, the complete open reading frames (ORF) of IL-11, ITLN and PNP5a genes were cloned and sequenced. The full-length cDNAs of the three genes contained an ORF of 597, 945 and 882 bp, encoding a polypeptide of 198, 314 and 293 amino acids, respectively. Phylogenetic analysis indicated that the three genes shared high homology to other bony fish species including Zebrafish. Interestingly, the ITLN gene of crucian carp lacked a 10 aa peptide that was found in the C-terminal of other fish species. A real-time RT-PCR assay was developed to quantitatively examine their tissue distribution. We found that IL-11, ITLN and PNP5a were expressed at low levels in all of the tissues examined. To monitor the response of these genes to CyHV-2 or Aeromonas hydrophila (A. hydrophila) infection, we determined the expression level of IL-11, ITLN and PNP5a at different time points after infection in kidney. Significant up-regulation of IL-11, ITLN and PNP5a was only observed 72 h post-CyHV-2 injection ($P < 0.01$), whereas significant up-regulation was observed as early as 6 h after infection with A. hydrophila ($P < 0.01$). Our results demonstrated that host innate immune response to CyHV-2, at least in which IL-11, ITLN and PNP5a were involved, was slow in comparison to that induced by A. hydrophila. It suggested that CyHV-2 might suppress host innate response during early infection. The lack of a C-terminal peptide of crucian carp ITLN gene implied a possible functional difference of this gene during evolution, which merit further investigation.

*Predictions for document #2*

```
+--------+-------------------------------+-----------------------------------------+--------------+-----+
|document|keyword                        |probability                              |probability_tf|label|
+--------+-------------------------------+-----------------------------------------+--------------+-----+
|24636855|aeromonas hydrophila           |[0.7883642648209124,0.21163573517908768] |false         |1    |
|24636855|carassius auratus gibelio      |[0.745505895329291,0.25449410467070904]  |true          |0    |
|24636855|crucian carp                   |[0.6998814237765938,0.3001185762234063]  |true          |0    |
|24636855|cyhv                           |[0.9460562614769064,0.05394373852309366] |false         |0    |
|24636855|cyprinid herpesvirus 2         |[0.745505895329291,0.25449410467070904]  |true          |1    |
|24636855|genes                          |[0.9528132100204569,0.04718678997954299] |false         |0    |
|24636855|il-11                          |[0.8355713769724151,0.16442862302758487] |false         |0    |
|24636855|interleukin 11                 |[0.6867481674413576,0.31325183255864236] |true          |1    |
|24636855|itln gene                      |[0.9570123210684232,0.042987678931576935]|false         |0    |
|24636855|purine nucleoside phosphorylase 5a|[0.6947078054929541,0.30529219450704603]|true       |1    |
|24636855|regulation                     |[0.9628005854012672,0.037199414598732834]|false         |0    |
+--------+-------------------------------+-----------------------------------------+--------------+-----+
```